# PREPRINT

# An empirical study on V2X radio coverage using leaky coaxial cables in road crash barriers

HagenUßler[a]*, Christian Setzefand[b], Oliver Michler[a]

[a]*TU Dresden, Hettnerstr. 3, Dresden, 01069, Germany*
[b]*MRK Media AG, An der Frauenkirche 12, Dresden, 01067, Germany*

**Abstract**

For current and future automated driving functions, the radio availability of broadband hybrid networking services (e.g. digital broadcasting, mobile radio, dedicated short range communication) is a prerequisite for continuous V2X information exchange. The supply focus for this is explicitly the road route with its lanes. The application of antenna-based solutions for such longitudinal radio cells with hybrid telematics services is expensive from the installation point of view and can only be adapted to new future telematics standards with great effort. A more suitable solution for such longitudinally shaped radio cells for road routes are leaky coaxial cables (LCX), which are already successfully used for tunnel solutions, for example. The paper discusses the installation and radio implementation of broadband LCX solutions (up to 6 GHz) in terms of simulation and surveying. The integration of the LCX into the crash barrier is favored due to low installation effort and easy upgradeability. An installation was realized on an automotive test fields, where preliminary empirical results for radio simulation and coverage were obtained. Based on the simulations and evaluation measurements, it can be shown that the propagated coverage approach is sustainable over all radiated services. Further solution approaches such as the direct insertion of LCX into the roadway and the derivation of vehicle location information are discussed in the outlook of the paper.

## 1. Introduction

In recent years, there has been growing academic and commercial interest in exploring applications using leaky coaxial cables (LCX). The conventional areas of application are mines and tunnels – see also Salem and Yousaf (2019), as these are long and narrow coverage areas and the radiation characteristics of LCX have good advantages here compared to point antenna solutions. Other advantages include the simple robust installation and the typically high frequency bandwidth of the LCX. Current research and development directions are also focused on wireless access systems with high frequencies and larger bandwidths, such as mobile radio 5G in the 3.5 GHz frequency band and Wi-Fi in the 6 GHz frequency band. For example, the suitability of LCX for potential V2X applications in tunnels is investigated and reported in Farahneh and Fernando (2019). Other research activities as reported in Blaszkiewicz et al. (2020) deal with the localization and position estimation of a mobile terminal along the cable. Prior investigations concerning the radiation characteristics of LCX have been carried out in Siddiqui et al. (2020) and focus on increasing the radiation efficiency by a method to periodically modulate the aperture of an LCX. In Matković and Šarolić (2019), a method for software-based simulation of signal coupling losses along slotted coaxial cables is described and demonstrated by measurement.

---

* Corresponding author. Tel.: +49-351-46336832; Fax: +49-351-46336782
  *E-mail address:* hagen.ussler@tu-dresden.de

With the entry of connected driving in the automotive sector, questions of communication signals availability as a basis for automated driving functions are significant. In the current state of the functional automation, this particularly concerns motorways and trunk roads. In addition to conventional tunnel solutions, the LCX can profitably be introduced into the road infrastructure. Integration into the typically existing crash barriers is particularly suitable, as this allows adequate and linear radio coverage with a high Quality of Service (QoS) in limited areas along the cable.

In this paper, the use of LCX as a broadband supply medium for hybrid telematics services (cf. Fig. 1) is discussed. This contributes to continuous radio coverage along roads and thus to the realization of V2X applications. Investigations were carried out for various selected frequencies and services. Methodically, computer-aided radio propagation simulations and measurements are conducted and discussed in a laboratory environment and a test field for connected and automated driving, where the LCX was installed along a road crash barrier for the first time.

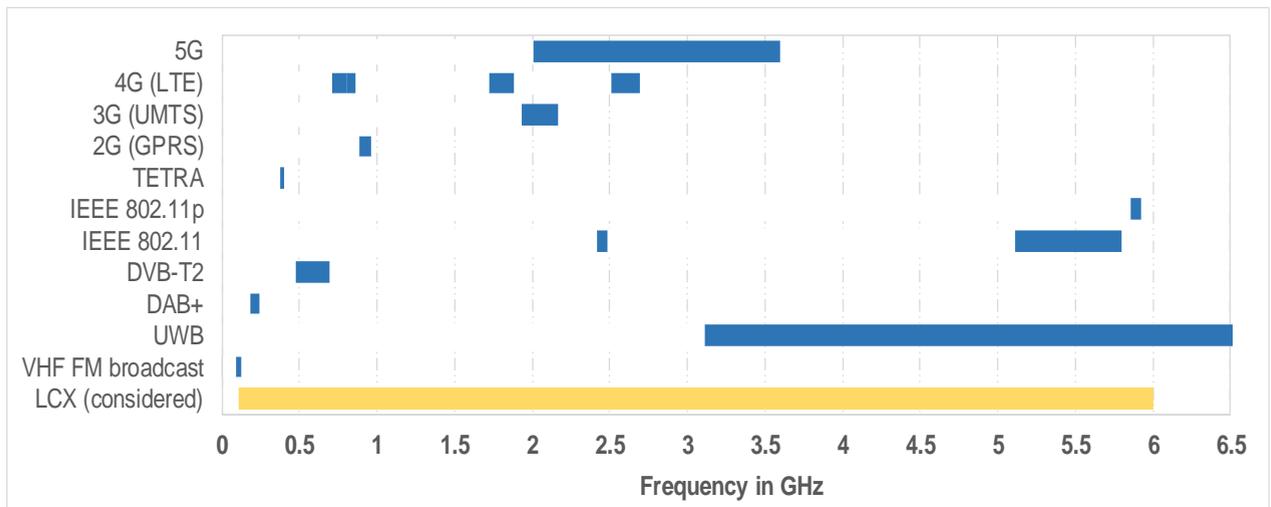

Fig. 1. Hybrid traffic telematics services and associated frequency ranges covered by LCX.

## 2. Leaky cable characteristics

LCX is a radiating coaxial cable, acting as an electric wave guiding coaxial cable and radiating antenna structure at the same time. It differs from an ordinary coaxial cable in that it has periodic slots along the outer conductor. This allows the electric energy of the used radio frequency (RF) to radiate in and out of the antenna as the electric wave propagates along its length. Fig. 2 shows the basic structure of a LCX. As investigated in Hassan et al. (2016), the shape, spacing and size of the slots are responsible for the loss, radio patterns and coupling properties. LCX are classified into two types in terms of their radio radiation mechanism, namely radiated mode and coupled mode.

As a basis to evaluate the LCX overall system performance, the link-budget can be considered according to Sesena et al. (2013) with

$$P_R = P_T - L_L - L_C - 10p \, log_{10}(d_{lat}) - L_{con} + G_R , \qquad (1)$$

where $P_R$ represents the received power in dBm, $P_T$ the transmit power in dBm, $L_L$ the LCX specific longitudinal loss in dB, $L_C$ the LCX specific coupling loss in dB, $d_{lat}$ the nearest lateral distance between the cable axis and the receiver antenna in m, p the loss exponent, $G_R$ the receiver antenna gain in dBi and $L_{con}$ the losses due to connectors, LCX feeding and antenna cables in dB.

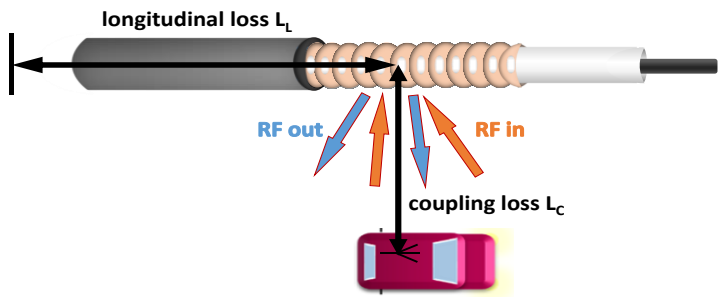

Fig. 2. Physical structure of a LCX and system loss correlations.

The basic equation relating the parameters of a leaky feeder system are $L_L$ and $L_C$. These LCX parameters form the basis for the simulations, measurements and coverage assessments in the following sections - see Wang et al. (2001).

$$L_L = \alpha \cdot d_{lon} , \qquad L_C = 10 \, log_{10}\left(\frac{P_{Tx}}{P_{Rx}}\right) \qquad (2)(3)$$

$L_L$ is the attenuation of the line between the base station and the point on the line nearest the mobile set and therefore equal to the LCX longitudinal distance, $d_{lon}$, in m, between these points multiplied by a frequency-specific attenuation factor, $\alpha$ in dB/m.

$L_C$ is the ratio of the power $P_{Tx}$ transmitted in the cable to the received power $P_{Rx}$ at the antenna of the mobile set. Usually, $L_C$ is given as a statistical value at which a defined percentage of all measured values has lower attenuation. Thus, $L_C$ is specified as a function of signal frequency, $f$, as 50 % and 95 % quantile at $d_{lat}$ = 2 m measured with a half-wavelength dipole antenna according to DIN EN 61196-4 (2004). Therefore, $G_R$ needs to be used relative to the dipole.

## 3. Simulative and experimental results

### 3.1. Radio propagation simulation

As a planning tool, radio propagation simulation allows reproducible investigation of propagation effects and calculation of radio coverage, providing a fast and efficient way to perform optimized infrastructure dimensioning for radio networks. Thus, this also serves as a basic method for evaluating radio propagation with LCX.

As indicated in Schwarzbach et al. (2020), radio propagation algorithms can be classified into empirical, semi-empirical, numerical and ray-based approaches. Since empirical models are derived from measurements, they are only valid for situations similar to those in which their database was obtained. Considering relevant transmissions as separate paths is an assumption that leads to ray-based propagation models. Compared to empirical models, they are more flexible if the propagation environments can be individually modelled. An environmental model includes the elements of the environment relevant for radio propagation, such as geometric and electrical properties (cf. Fig. 3-a). Examples of typical application areas and descriptions of different radio propagation approaches can be found in aircraft cabin in Ringel et al. (2013), public bus in Michler et al. (2015) or parking garage in Jung et al. (2020).

For the radio propagation simulation performed in this work, the software WinProp from Altair Engineering, Inc. (2021) is used. The investigations are carried out using a CAD model of the test field for automated driving at the University of Applied Sciences (HTW) Dresden. For technical and functional reasons such as frequency range and attenuation characteristics, ½" LCX with radiating mode of RFS company are considered for both, radio simulation and measurements– see RFS (2020). The LCX was parameterized with frequency dependent coupling loss $L_C$ and cable loss $L_L$ and modeled as a large number of antennas in a row with specific discretization interval, depending on the propagation model used. For the simulation of radio propagation, the *Dominant Path Model* (DPM) according to Wahl and Wolfle (2006) was used to consider the most relevant path in terms of signal energy between transmitting antenna on the LCX line and receiving antenna. In addition, a simulation with *Smallest Path Loss* model (SPL) was performed, which considers direct signal path and a small antenna discretization interval. The receiver plane is modeled as an array of individual receive spots with 1x1 m resolution to capture and represent spatial distributions.

The simulation was performed for different frequencies in the range between 0.1 to 5.9 GHz to cover several services used in V2X communications or for general information purposes in transportation, specifically radio broadcast and Traffic Message Service (TMC) in the 0.1 GHz band. Digital Audio Broadcast (DAB+) (0.2 GHz), 2G mobile communication (0.9 GHz), 4G mobile communication (1.8 GHz), 5G mobile communication (3.8 GHz), IEEE 802.11 Wi-Fi technology (2.4 GHz) and IEEE 802.11p for V2X-Technology (5.9 GHz). Fig. 3-b shows simulation results with $f$ = 5.9 GHz and 100 m LCX attached to the crash barrier, starting at the right side, along the curve and upper side. Given $P_T$ = 18 dBm, receive powers greater than $P_{Rx}$ = -85 dBm are simulated over the total area of the test field (50 m · 100 m). Simulation results in the lower frequency bands show particularly better coverage results as expected due to distance and frequency dependent free space propagation. With known receiver sensitivity of the receiver hardware, position-based conclusions can thus be drawn about the reception quality. Therefore, with correct calibration, the radio planning simulation provides a way to efficiently evaluate the QoS and radio coverage.

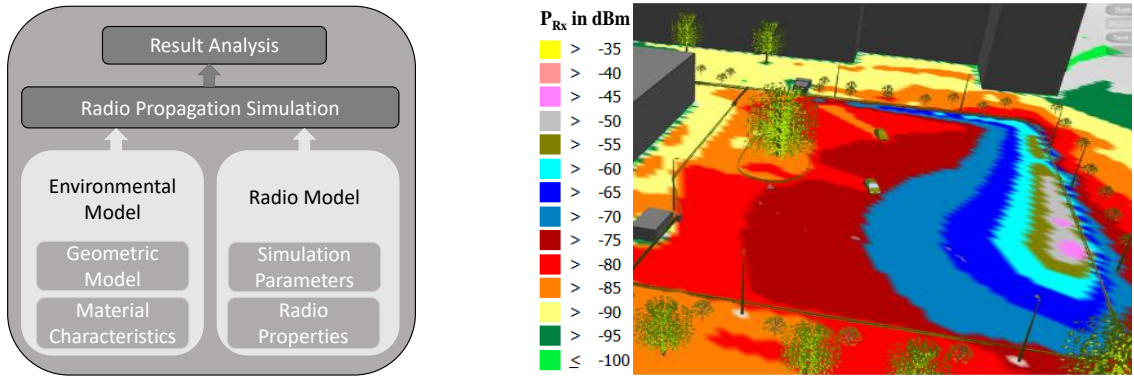

Fig. 3. (a) Components of a radio propagation simulation; (b) Modeling of test field environment for automated driving, including simulation results for $f$ = 5.9 GHz.

### 3.2. Anechoic chamber measurements

The analysis of the radiation characteristics of the LCX in the crash barrier is carried out in an anechoic chamber in order to exclude the influence of external disturbance variables on the measurement results.

Conventional LCX installations, e.g. in tunnels, require a specific, fixed distance between the cable and the installation surface in order to avoid interference, especially with other metallic bodies and surfaces. Installation along metallic crash barriers differs from this requirement due to the limited LCX mounting options. Therefore, especially with regard to traffic telematics applications along roads, potential signal interference effects have to be evaluated. To fix the LCX, holes were drilled in the crash barrier and metal brackets were fixed by means of screw connections. To investigate the influence of the curvature of the crash barriers surface and the installation position of the cable, the mounts were additionally installed at central top, central middle and central down.

According to Fig. 4-a, a section of the assembled crash barrier was measured, especially with regard to the coupling losses. Continuous wave (CW) measurements were carried out systematically at longitudinal distances of 50 centimeters along the cable and for lateral distances of one, two, three and four meters. This resulted in 20 measurement points per frequency examined. Antenna and LCX are mounted at the same height (approx. 60 cm). Due to the large frequency range investigated, from 0.1 GHz to 6 GHz, three different omnidirectional antennas with defined antenna gain were used in frequency ranges to be investigated and controlled via an RF-switch. For each measuring point, 100 measurements were recorded for the frequency range supported by the antenna used. The LCX was terminated by means of a 50 Ohm resistor. The block diagram of the measurement setup is shown in Fig. 4-b.

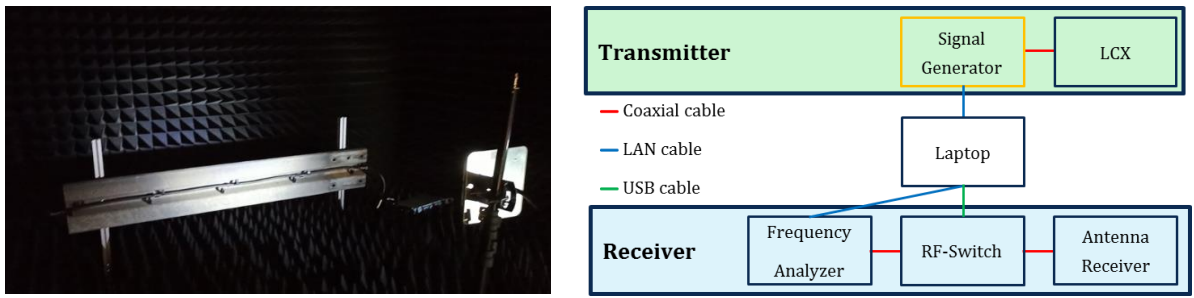

Fig. 4. (a) LCX measurement environment in the anechoic chamber with crash barrier; (b) Block diagram of the measurement setup.

In order to be able to assess the influence of the crash barrier on the signal propagation, CW measurements were first carried out with a free-hanging LCX fixed to a wooden frame. The frequency-specific results for the respective installation positions are shown in Fig. 5. The cable and connection losses and the respective frequency-specific antenna gains of the measurement equipment used, were taken into account in the results. The cable and connection losses, $L_{con}$, were determined by measurements. The frequency-specific antenna gains were taken from the data sheets of the measuring antennas used.

The results of the laboratory measurements show that the free-hanging and crash barrier installation produce measurement results that differ greatly in some cases and that the LCX installation environment thus has a demonstrable influence on signal propagation. The largest measured difference between the installation in the crash barrier and the free-hanging installation was 29 dB for $f$ = 0.9 GHz. For installation central top, $P_R$ = -38 dBm was measured, while -67 dBm was determined for free-hanging LCX. Furthermore, it could be deduced that the effect of the LCX installation environment is frequency-dependent. For $f$ < 0.3 GHz, the measurements with the free-hanging installation show higher signal power levels than those with the crash barrier. At 0.9 GHz, the largest signal amplification with crash barrier installation was observed for all frequencies investigated.

The examination leads to the result that a signal amplifying effect for higher frequency signals is achieved by the crash barrier. In particular, this is due to the signal-reflecting effect of the metallic surface. As a result, the radially coupled-out signal components of the LCX are reflected at the metal notch in the direction of the receiving antenna and interfere. Therefore, both signal power amplification and attenuation can occur. The antennas changed radiation characteristics, when influenced by metallic objects in the immediate vicinity, is also described in Volakis (2019). Here, the effective area and the diameter of the antenna aperture (dimension of the crash barrier) must be larger, the longer the signal wavelength, $\lambda$. The measurement results show an increase in signal attenuation for frequencies in the range of 0.1 GHz to 0.2 GHz, resulting in lower signal reception power. It is therefore assumed that the LCX used in the anechoic chamber with a $d_{lon}$ = 2 m is too short for the correspondingly longer signal wavelengths since $d_{lon} > 10 \cdot \lambda$ is also recommended in DIN EN 61196-4 (2004). Effects in the upper kHz range are therefore to be further investigated with longer LCX. However, it is presumed that similar signal amplifying effects will be observed here.

Comparing the installation positions of the LCX along the crash barrier, the central middle position was assumed to be the base configuration. The measurements show a frequency-dependent power gain of at least 4 dB compared with free-hanging installation. Furthermore, frequency-dependent deviations in the received power levels, with significantly worse measured values being observed for the central top position in the frequency range 5.0 to 6.0 GHz, for example. The result analysis does not provide a general statement on the best crash barrier installation position.

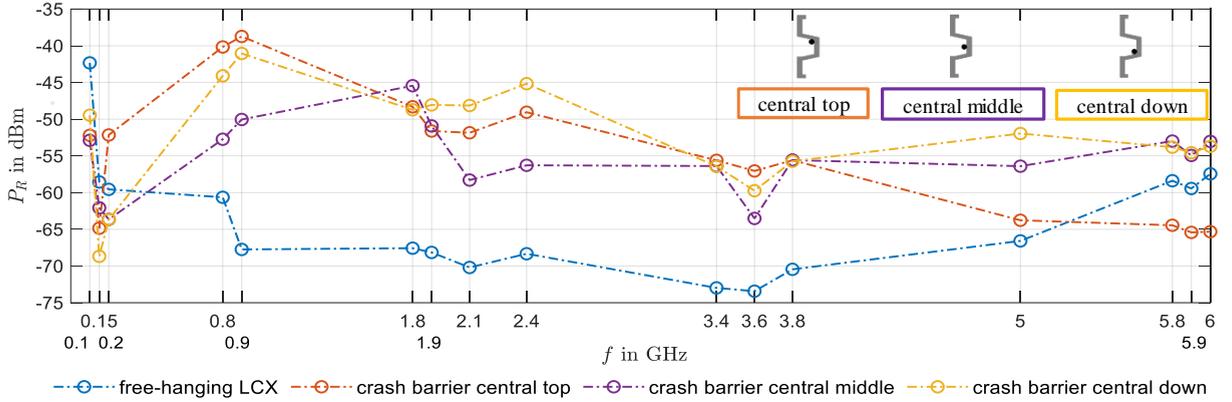

Fig. 5. Received power measurements as a function of the investigated frequency range for different crash barrier installation positions of the LCX within the laboratory environment.

### 3.3. Field test set up and measurements

For the field test investigations, LCX were integrated within a section of crash barriers in the testbed for automated and connected driving at HTW Dresden, which was already introduced for simulation purposes in sec. 3.1. At first, CW measurements were performed for all frequencies considered in order to investigate the radio coverage under real conditions and over longer $d_{lat}$ up to 32 m. The length of the LCX initially was $d_{lon} = 15$ m, measurements were performed at 1 m intervals in the longitudinal direction and $d_{lat} = 2^n$ m with $n \in 0...5$ in the lateral direction. 100 measurements were recorded per measurement point and frequency. The field test measurement setup according to Fig. 4-b and the measurement vehicle, used for telematics service measurements, is shown in Fig. 6-a.

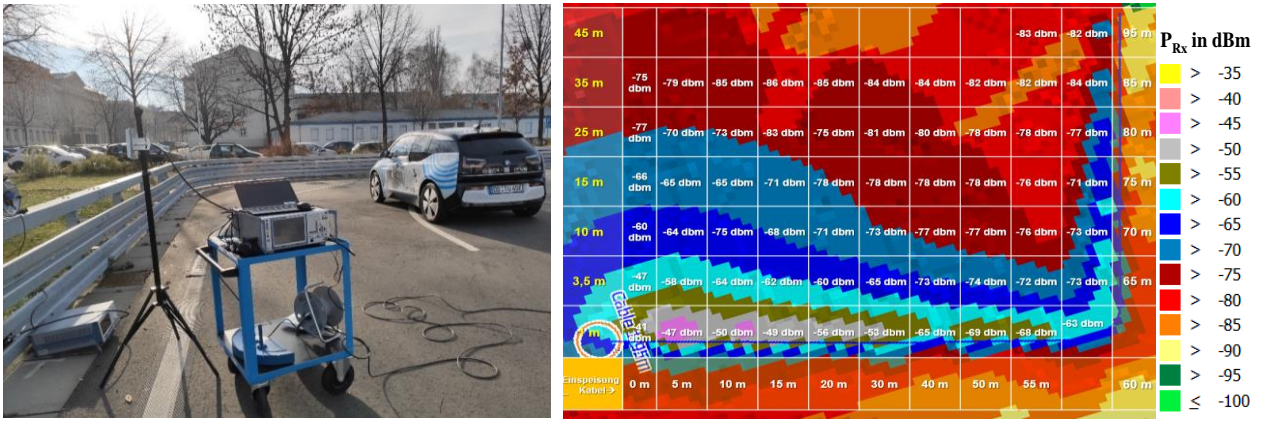

Fig. 6. (a) Measurement setup in test field with crash barriers; (b) CW measurements and simulation results for $f = 5.9$ GHz in the test field.

The received measured power at the receiving antenna, $P_R$, with $P_T = 18$ dBm are shown in Fig. 7 (top row) for selected frequencies, classified in steps of 10 dBm. The minimum received power over all frequencies considered was found to be -92 dBm at $f = 0.2$ GHz. Similar to the measurement results in the anechoic chamber, signals with frequencies above at least 0.2 GHz can be received with a higher signal level. As the signal frequency then increases, the signal level drops continuously, as expected. Due to the short length of the measured LCX as well as the variance of the measured values, no significant longitudinal attenuation could be shown in this measurement setup. However, between the discrete measurement points at same $d_{lat}$, large power level differences were observed in some cases. With LCX, these occur as fading notches due to the radiation characteristics caused by interference and can be observed every half $\lambda$ according to Weber et al. (2010).

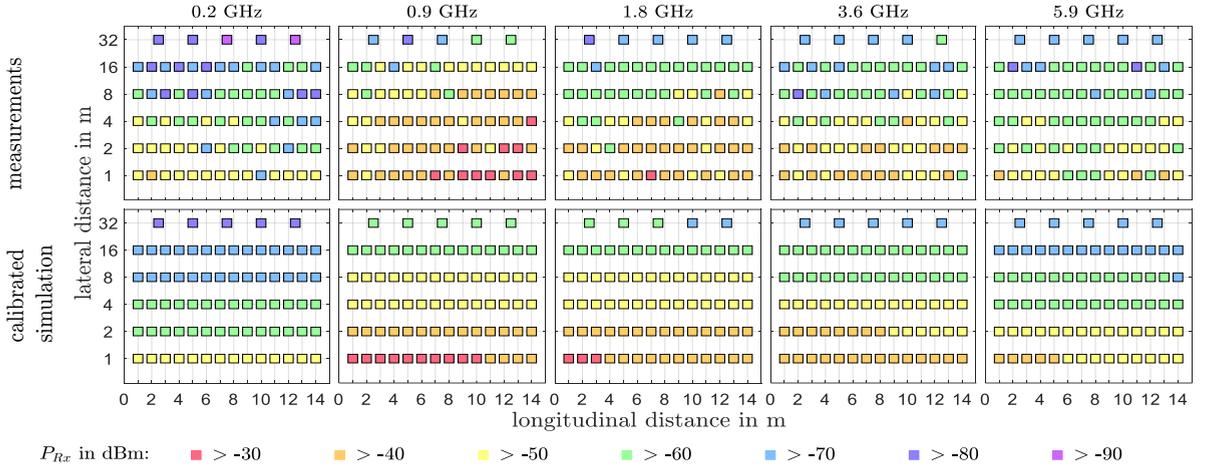

Fig. 7. Measured and simulated received power level at the antenna for selected frequencies and positions regarding LCX in the test field.

A comparison of the power levels simulated with the radio planning software showed that it provides conservative results. On the one hand, the influence of the crash barrier is not sufficiently taken into account in the propagation model as it cannot fully represent the real level of detail; on the other hand, the $L_C$ specified in the LCX data sheet and used in the propagation model is given with a tolerance of 10 dB – see RFS (2020).

To calibrate the radio propagation model, the given coupling losses for the LCX attached to the crash barrier were recalculated from the measured level values at the test site using the Eq. (1), (2), (3). According to DIN EN 61196-4 (2004) measurements at $d_{lat}$ = 2 m were considered. The radio planning simulation with calibrated coupling loss is shown with respect to the measurement points in Fig.7 (bottom row) for selected frequencies. In Tab. 1-a, the mean power deviation of the simulation results from the measured values is given for all longitudinal positions with respect to $d_{lat}$ and frequency. Mean deviation up to 7 dB despite calibration was determined. This is due to the relatively large scatter of the level measurement values in the longitudinal direction due to the described fading notches. These are not modeled in this level of detail in the radio propagation simulation respectively the used radio propagation model.

For frequencies of 5.9 GHz, relevant for V2X applications, measurements were performed with a LCX of 100 m length, which also extends along the curve area of the test field. The CW measurement results are shown together with the calibrated radio planning results in Fig. 6-b. Again, only minor deviations between the simulation model and the CW measured values were observed. In general, the received power does not fall below $P_R$ = -86 dBm in the entire test field and therefore full radio coverage and power level is provided, which is necessary for V2X services.

Table 1. (a) mean error between measurement and calibrated simulation in dB; (b) RSSI of radio module for ITS-G5 service at 5.9 GHz in dBm.

| $f$ in GHz | | 0.1 | 0.2 | 0.9 | 1.8 | 2.4 | 3.6 | 5.9 |
|---|---|---|---|---|---|---|---|---|
| $d_{lat}$ in m | 1 | 0.3 | 1.6 | -2.4 | -5.8 | -4.9 | -5.8 | -5.6 |
| | 2 | -0.4 | -0.9 | 0.5 | -1.1 | 0.2 | -1.2 | -2.3 |
| | 4 | 2.6 | 1.9 | 5.7 | -1.2 | 1.9 | -2.0 | 0.8 |
| | 8 | 1.6 | -0.4 | 7.2 | -3.4 | 5.3 | -5.7 | 2.6 |
| | 16 | -4.0 | 2.5 | 4.4 | 0.6 | 4.1 | -1.7 | 3.7 |
| | 32 | -5.2 | -5.7 | -3.0 | -7.4 | -2.8 | 0.4 | 2.8 |

| $d_{lon}$ in m | | 2 | 7 | 12 |
|---|---|---|---|---|
| $d_{lat}$ in m | 2 | -37 | -48 | -37 |
| | 8 | -44 | -47 | -54 |
| | 16 | -54 | -51 | -54 |
| | 32 | -62 | -54 | -55 |
| | 50 | -68 | -61 | -63 |

In addition to the CW measurements of the LCX in the crash barrier, measurements of selected communication services, in particular for V2X applications, were performed in a final step. The reception of the now modulated signals depends on the receiver sensitivity of the hardware. V2X services based on IEEE 802.11p at 5.9 GHz were examined in detail. The ETSI ITS-G5 standard along with Cohda Wireless MK5 module was used, reporting a receiver signal strength indicator (RSSI) – see De Haaij et al. (2017). The module was configured with $P_T$ = 18 dBm. The measurements recorded with the research vehicle fundamentally showed that service reception is possible in the entire test field up to $d_{lat}$ = 50 m, regarding RSSI. The receiver sensitivity of -95 dBm was never undercut (cf. Tab.1-b). Packet error rate was observed to be 0 % for each measuring point with 100 ITS-G5 messages sent each. These empirical results show that LCX in crash barriers are suitable for radiating V2X communication services along roads.

## 4. Summary and outlook

The results of the simulative, laboratory and field test-based investigations empirically prove that LCX are suitable infrastructure tools for antenna installations along roads and highways with equipped crash barriers and can potentially compensate shortcomings of conventional antenna deployment. The investigations show that direct mounting in crash barriers results in a signal amplification effect for higher frequency signals compared to a free-hanging LCX. Advantages are simple installation, robust operation but also optimally tailored coverage area of several lanes. Furthermore, due to the broadband nature of the LCX, many telematics services can be transmitted and broadcast in parallel, which is very cost-effective and important for actual and future V2X services. Currently, extended investigations are being conducted on the use of LCX along crash barriers for vehicle localization in the laboratory as well as in the test field. These are Time Difference of Arrival (TDoA) investigations in the 6 GHz Ultra-wideband (UWB) with 500 MHz bandwidth. Future investigations of LCX for V2X applications could include placing the LCX directly in the roadway, where there are no crash barriers along the road.

## References

Altair Engineering, Inc.: Altair Feko™-Suite. Altair WinProp 2020. https://www.altair.de/feko-applications/

Blaszkiewicz, O., Sadowski, J., Stefanski, J., 2020. Position Estimation in Corridors Along the Coupled Mode of Radiating Cables. Sensors 2020, 20, 5064. https://doi.org/10.3390/s20185064

De Haaij, D., Strauss, U., Sloman, M., 2017. Cohda Wireless MK5 OBU Specification, Reference: CWD-P0052-OBU-SPEC-WW01-186, V. 1.6

DIN EN 61196-4, 2004. Coaxial communication cables, Part 4: Sectional specification for radiating cables (IEC 61196-4:2004)

Farahneh, H., Fernando, X., 2019. The Leaky Feeder, a Reliable Medium for Vehicle to Infrastructure Communications. Applied System Innovation. 2019; 2(4):36. https://doi.org/10.3390/asi2040036

Hassan, N., Fernando, X., Farjow, W., 2016. Optimization of Leaky Feeder Slot Spacing for Better Beam Forming in Mines and Tunnels. International Journal of Communications, Network and System Sciences, vol. 9 no.4, pp. 1–8. 2016.

Jung, A., Schwarzbach, P., Michler, O., 2020. Future Parking Applications: Wireless Sensor Network Positioning for Highly Automated In-House Parking,  Proceedings of the 17th Intern. Conf. ICINCO, Paris, 710-717.

Matković, A., Šarolić, A., 2019. Slot Antenna in a Coaxial Cable Shield – Coupling Loss Computational Analysis. *2019 European Microwave Conference in Central Europe (EuMCE)*, pp. 78-81.

Michler, O., Weber, R., Förster, G., 2015. Model-based and empirical performance analyses for passenger positioning algorithms in a specific bus cabin environment, Proceedings of the 4th Intern. Conf. on MT-ITS, Budapest, Hungary, 200-208.

RFS: Product Datasheet, RCF12-50JFN, 1/2" RADIAFLEX® RCF Cable. Rev: A. Rev Date: 30 Oct 2020.

Ringel, J., Klipphahn, S., Michler, O., 2013: Simulation of Wave Propagation for Radio and Positioning Planning inside Aircraft Cabins. Proceedings of the 3rd Intern. Conf. on MT-ITS, Dresden, 243-253.

Salem, A., Yousaf, S. S. N., 2019. An investigation of wideband MIMO channel characteristics in rectangular tunnel, International Transaction Journal of Engineering, and Applied Sciences, vol. 10, no. 15, 1-15.

Sesena, J., Aragón-Zavala, A., Zaldivar-Huerta, I., Castanon, G, (2013). Indoor propagation modeling for radiating cable systems in the frequency range of 900-2500MHz. Progress In Electromagnetics Research B. 47(1). 241-262. 10.2528/PIERB12102314.

Schwarzbach, P.; Engelbrecht, J.; Michler, A.; Schultz, M.; Michler, O., 2020. Evaluation of Technology-Supported Distance Measuring to Ensure Safe Aircraft Boarding during COVID-19 Pandemic. Sustainability 2020, 12, 8724. https://doi.org/10.3390/su12208724

Siddiqui, Z., Sonkki, M., Tuhkala, M., Myllymäki, S., 2020. Periodically Slotted Coupled Mode Leaky Coaxial Cable With Enhanced Radiation Performance. In IEEE Transactions on Antennas and Propagation, vol. 68, no. 11, pp. 7595-7600, Nov. 2020, doi: 10.1109/TAP.2020.2990478.

Volakis, J. L., 2019. Antenna engineering handbook. Fifth Edition. McGraw Hill Education, New York. 2019

Wahl, R., Wolfle, G., 2006. Combined urban and indoor network planning using the dominant path propagation model, First European Conference on Antennas and Propagation, Nice, 2006, pp. 1-6.

Wang, J. H., Mei, K. K., 2001. Theory and Analysis of Leaky Coaxial Cables with Periodic Slots. In IEEE Trans. on Antennas and propagation, vol. 49, no. 12, pp. 1723-1732.

Weber, M., Birkel, U., Collmann, R., Engelbrecht, J., 2010. Comparison of various methods for indoor RF fingerprinting using leaky feeder cable. 7th Workshop on Positioning, Navigation and Communication, 2010, pp. 291-298, doi: 10.1109/WPNC.2010.5653786.